\documentstyle[preprint,aps]{revtex}
\draft
\begin{document}
\newcommand{\be}{\begin{equation}}
\newcommand{\ee}{\end{equation}}
\title{Spatial Coherence of Tunneling in Double Wells.}
\author{L. S. Levitov$^{a,b}$ and A. V. Shytov$^b$}
\address{(a) Massachusetts Institute of Technology,
12-112,
77 Massachusetts Ave., Cambridge, MA 02139}
\address{(b) L. D. Landau Institute for Theoretical Physics, 2, Kosygin st.,
Moscow, 117334, Russia}
\maketitle
\begin{abstract}
Tunneling between two 2D electron gases in a
weak magnetic field is of resonance character, and involves a
long lifetime excitonic state of an electron and hole {\it
uniformly spread} over cyclotron orbits. The
tunneling gap is linear in the field, in agreement with the
experiment, and is anomalously sensitive to the electron density
mismatch in the wells. The spatial coherence of tunneling along
the orbit can be probed by magnetic field parallel to the plane,
which produces an Aharonov-Bohm phase of the tunneling amplitude,
and leads to an oscillatory field dependence of the current.
  \end{abstract}
%\narrowtext
%\twocolumn
%\noindent{\it   Introduction.}\hskip4mm
Electrons in GaAs quantum wells have very high mobility, and at
low temperature form an almost ideal Fermi liquid. Strong Coulomb
interaction and sensitivity to external magnetic fields make the
physics of this system rich and interesting. Recently, several
phenomena have been discovered in tunneling experiments in
quantum wells, including: a  tunneling gap induced by a magnetic
field\cite{GapGeneral}; resonance peaks of conductivity near zero
bias\cite{B=0peak}; and excitonic effects\cite{excitonic}. Typically,
one has two wells containing 2D Fermi liquids with an   electron
density of the order of $10^{11}{\rm cm}^{-2}$, separated by an
oxide barrier of few tens   nanometers thick.
It is characteristic that the barrier thickness is
very uniform, so that tunneling is  coherent  in  lateral  dimensions.
This  results  in  conservation  of both momentum and energy.
Basically, at a
zero magnetic field, only tunneling between identical
plane wave states can occur, which restricts the phase space of  final
states  and  leads  to an  unusual $I-V$ curve with a sharp peak near
zero bias\cite{B=0peak}. Finite width of the peak is  determined
by elastic  scattering.  One  can  probe  spatial  coherence  by
applying a magnetic   field  parallel  to  the  barrier.  Having
practically no effect on  the  dynamics  in  the  plane, the  vector
potential  of  the  parallel field shifts electron momenta in one plane,
which creates a mismatch of  the  Fermi  surfaces  of  different
planes. The effect on the  tunneling rate can be easily accounted for
by the free electron picture\cite{B=0peak}.

A magnetic field applied perpendicular  to  the  plane  makes  the
situation  more  interesting, especially in high fields, when
the system is in the Quantum Hall  state\cite{GapGeneral}.
Compared  to  the  $I-V$ curve at zero field, the current peak is
shifted away from zero bias to some finite  voltage.  Also,  the
peak broadens as the field increases. The current is almost entirely
suppressed  below the lower  edge  of  the peak, called a ``tunneling
gap.''
  %Further up in voltage, above the peak, there
%are weak features corresponding to tunneling between different
%Landau levels.
   Recently the tunneling  gap  has  been  intensively  studied
because  it  is  believed  that  it  can be used to probe the QH
state. The gap depends on  the  field  linearly  at  weak  field
($\nu\gg1$)\cite{WeakFieldGap}, and saturates at higher field
($\nu\simeq 1$) \cite{GapGeneral}. To summarize, one can
say that the $I-V$ curve can be interpreted in terms of a
resonant tunneling mechanism, involving some intermediate state
with the lifetime given by inverse peak width, and of the work
needed to create this state corresponding to the gap.

The gap at high field is quite well
understood~\cite{GapAtNu=1/2,HighField,HighFieldExcitonic}: its
energy scale is of the order of $e^2/\epsilon a$, where
$a=n^{-1/2}$ is the interparticle distance, and $\epsilon$ is
the dielectric constant. Until recently, the low field gap has
received less attention. The only available theory is by
Aleiner, Baranger, and Glazman~\cite{Glazman}, who developed a
hydrodynamical picture by treating the system as an ideal
compressible conducting liquid. They wrote down classical
electrodynamics equations in terms of charge and current
densities, and derived the tunneling gap
$\Delta=(\hbar\omega_c/\nu)\ln(\nu e^2/\epsilon\hbar v_F)$. At
constant electron density, they predict quadratic dependence of
the gap on the magnetic field, which is different from the
linear dependence found experimentally~\cite{WeakFieldGap}.

The reason for the disagreement, in our opinion, is that in a
clean metal, such that $\omega_c\tau\gg1$, one can use classical
electrodynamics only on a scale much bigger than the cyclotron
radius $R_c=v_F/\omega_c$. However, we will argue that the
important scale of the problem is of the order of $R_c$, and
thus one has to have a Fermi liquid in magnetic field. On this
scale, the state formed at tunneling has a non-trivial spatial
structure, which makes the physics very different from that of
the gap in high fields. Also, we will find that at weak field a
large simplification occurs, because the problem is
semiclassical, and one can use the classical Fermi liquid
equation to describe the dynamics in terms of Fermi surface
fluctuations.

Another point is that in two dimensions the energy and  momentum
conservation prescribes that all quantum numbers of the final and
initial tunneling states  coincide.  In a  magnetic field, this implies that
the radius of an  electron orbit as well as the guide center of  the
orbit  are conserved at tunneling. In a weak field, this results
in spatial coherence of tunneling over a large  distance  of  the
order of $R_c$, and leads to interesting effects.

\noindent{\it Summary of results.}\hskip4mm
Our goal is to explain the linear field dependence of the gap in
a weak magnetic field, and to propose experiments that will
reveal spatial structure of the intermediate tunneling state.
Although at the end we are going to do a rigorous Fermi liquid
calculation, it is instructive to begin with a semiclassical
picture of electron states localized near classical cyclotron
orbits and weakly interacting with each other. At tunneling, an
electron hops from the orbit of radius $R_c$ in one layer to the
identical orbit in the other layer, and leaves a hole on the
first orbit. This creates an electrostatic configuration of two
oppositely charged rings of radius $R_c$ separated by the
barrier of width $d\ll R_c$. The energy of this charge
distribution is
  \begin{equation}\label{gap}
\Delta={e^2\over\epsilon \pi R_c}\ln{d\over l_B}\ ,
\end{equation}
   where $\epsilon$ is the dielectric constant, and the magnetic
length $l_B=\sqrt{\hbar c/e B}$ characterizes the ring's
``thickness.'' Basically, we are saying that, in order to
transfer an electron, one has to charge the ``two ring
capacitor,'' and its charging energy $e^2/2C$ constitutes the
tunneling gap. The result (\ref{gap}) holds for $l_B\le d$, i.e.,
fields which are not too low, and fall in the experimental
range~\cite{WeakFieldGap}. The gap $\Delta$ dependence on
magnetic field is nearly linear, since the log term is roughly
constant. The dependence on the barrier width is in agreement
with the excitonic picture~\cite{excitonic,HighFieldExcitonic}.
By the order of magnitude the gap (\ref{gap}) agrees
with the experiment~\cite{WeakFieldGap}, which raises the
question of why there is no gap suppression due to Coulomb
screening. Basically, the reason is that he charge istribution is
localized in a very thin ring, of the thickness $l_B$ comparable
to the screening length, which makes the screeing effects not too
dramatic: they simply change the log in Eq.(\ref{gap}) by a
constant of the order of one.

That the tunneling is coherent along the cyclotron orbit can be
easily verified by applying a magnetic field parallel to the
barrier, in addition to the perpendicular field. The parallel
field flux ``captured'' between electron and hole trajectories
will give an Aharonov-Bohm phase to the tunneling amplitude, and
make it proportional to
  \be\label{ABphase}
\int\!\int\,
e^{{ie\over c\hbar}R_cdB_\parallel(\cos\theta_1-\cos\theta_2)}
d\theta_1d\theta_2\ .
  \ee
Thus the current will
oscillate as a square of the Bessel function:
\begin{equation}\label{Bparallel}
I(V)=J_0^2\left({p_Fd\over\hbar B_\perp}B_\parallel\right)\
I(V)_{B_\parallel=0}\ .
\end{equation}
   For an ideal  two dimensional system, the current    dependence
on $B_\parallel$ factors out because a parallel  field  does  not
affect  the  motion  in  plane.  For real wells of finite
width,  the  factorization  (\ref{Bparallel})  should still be a
good approximation at small $B_\perp$, when the cyclotron radius
is big compared to the well width. However, since the
parallel  field  will  squeeze  the  states  in  the  wells,  and
effectively  increase the barrier width, the tunneling rate may
aquire an additional non-oscillatory suppression factor.

Also, it is clear from what has been said, that the gap will be
very sensitive to any asymmetry between the wells. For example,
if there is a small mismatch $\delta n$ of densities in the wells,
the radii of the electron and hole orbits will become different,
$\delta R_c\simeq(R_c/2n)\delta n$,
which will reduce mutual capacitance of the orbits, and thus
raise the electrostatic energy. The gap will be enhanced by the order of its
magnitude at
$\delta R_c\simeq l_B$, which corresponds to the density mismatch
$\delta n/n\simeq l_B/R_c$ anomalously small at weak field.

\noindent{\it  Fermi liquid calculation.}\hskip4mm
To study the screening effects, we will use the Landau equation
   \begin{equation}\label{LandauFermi}
\left(\partial_t-D(1+\hat F)\right)\delta n({\bf p},{\bf r},t)=0\ ,
  \end{equation}
where $D={\bf v}_p\cdot\nabla_r+\omega_c m {\bf
v}_p\times\nabla_p$ contains the Lorentz force term. In
Eq.(\ref{LandauFermi}) we have omitted a  collision integral, which
is legitimate in the semiclassical limit $\hbar\omega_c\ll E_F$.
We will find a stationary solution that fully accounts for the
screening of the electron and hole charges. The resulting
distribution consists of two parts: singular and smooth. The
singular distribution is localized near the electron and hole
cyclotron orbits, within a short distance of the order of
magnetic length. The smooth part is spread over a distance
larger than the cyclotron radius. It is worthwhile to make a
comparison with the standard picture of a particle in a Fermi
liquid, which consists of a quasiparticle excitation combined
with a background density fluctuation, and to draw a relation
with the singular and smooth densities.

For simplicity, let us assume that $\hat F$ corresponds to a
pure density-density interaction:
 \be\label{interaction}
(\hat F\delta n)_\alpha ({\bf p},{\bf r})=
\sum_{r',p',\beta}U_{\alpha\beta}(r-r')\delta n_\beta({\bf p'},{\bf r'})\ ,
\end{equation}
   where $\alpha,\beta=R,L$ label wells, and
$U_{\alpha\beta}(r)$ is the Coulomb potential. It will be
straightforward to modify the calculation for a more general
Fermi liquid interaction. One can write $\delta n({\bf p},{\bf
r})$ corresponding to the electron and hole being uniformly
spread over cyclotron orbits:
   \begin{equation}\label{e+h}
\delta n^{(0)}_{R(L)}({\bf p},{\bf r})=\pm  \int\limits_0^T{dt\over T}\,
\delta({\bf p}-{\bf p}(t))\,
\delta({\bf r}-{\bf r}(t))\ ,
%\left(\begin{array}{c}1\\ -1\end{array}\right)\ ,
\end{equation}
  where ${\bf r}(t)$, ${\bf p}(t)$ is the classical circular
trajectory with $|{\bf p}|=p_F$, and with the period
$T=2\pi/\omega_c$. Semiclassically, the density (\ref{e+h})
corresponds to a state of the lowest Landau level available for
tunneling. At tunneling, initially, the states (\ref{e+h}) are
created, and then, over the   time scale $\sim\omega_c^{-1}$, they
relax to a stationary state, whose energy gives the tunneling
gap.

In order to find the system response to the appearance of the
electron and hole, we have to solve the equation:
   \begin{equation}\label{actual_n}
\left(\partial_t-D(1+\hat F)\right)\delta n=\delta(t)\ \delta n^{(0)}\ .
  \end{equation}
Formally, the  solution of Eq.(\ref{actual_n}) can be written as a ``ladder'':
   \be\label{ladder}
\delta n=\left(G+GD\hat FG+GD\hat FGD\hat FG+...\right)\delta n^{(0)}\ ,
  \ee
where $G=(\partial_t-D)^{-1}$ is readily evaluated in the   Fourier
representation:
  \be \label{inverse_operator}
G({\bf k},{\bf p},{\bf p'}) = \sum_{n}
\frac{e^{in(\theta_p- \theta_{p'})
+ i{\bf k}\times({\bf p} - {\bf p'})/m\omega_c}}{2\pi i(n\omega_c-\omega)}\ .
\ee
   The density (\ref{e+h}) satisfies the equation
(\ref{LandauFermi}) with $\hat F=0$. The first term of the
ladder (\ref{ladder}) gives $\delta n^{(0)}({\bf p},{\bf r})\,\theta(t)$, the
singular part of the density. The rest is the smooth ``background''
density.

In terms of $\delta n$ the gap is given by
  \be\label{GapTime}
\Delta=\sum_{p,r,\alpha,p',r',\beta}
\delta n^{(0)}_\alpha({\bf p},{\bf r}) \hat  F^{\alpha\beta}
\delta n_{\beta}({\bf p}',{\bf r}',t\to\infty)\ .
  \ee
By summing up  the ladder (\ref{ladder}) in a standard Fermi  liquid
fashion, one gets
  \begin{eqnarray}
\label{GapFrequency}
\Delta=\sum_{k,p,p'} \delta n^{(0)}_{\bf -k,\alpha}({\bf p})
\hat {\cal F}^{\alpha,\beta}_{\omega,{\bf k}}
\delta n^{(0)}_{\bf k,\beta}({\bf p}')_{\omega\to0}\ ,\\
(1 + \hat F_k ) \hat{\cal F}_{\omega,{\bf k}}= \hat F_k
\left(
     1 + \omega \sum_n \frac{J_{n}^2(kR_c)}{\omega - n\omega_c}
                   \hat{\cal F}_{\omega,{\bf k}}
\right)\ ,
  \end{eqnarray}
Here  $J_n$  are Bessel functions. At $\omega\to0$ it turns into
$
\hat{\cal F}_{k}=
\hat  F_k\left(1+\hat F_k(1-J_0^2(kR_c))\right)^{-1}
$.
By plugging it in Eq.(\ref{GapFrequency}) together with the
Fourier transform of density,
$\int d^2{\bf p} \delta n^{(0)}_{\bf k,\,R(L)}({\bf p})=\,\pm\,J_0(kR_c)$,
we  finally  get
  \be\label{Gap}
\Delta = \int\limits_0^{\infty} \frac{J_0^2(kR_c) e^2
U_k}{1 + e^2\nu U_k\left(1-J_0^2(kR_c)\right)}\,
\frac{k\,dk}{2\pi}
  \ee
where $U_k=2\pi(1-e^{-kd})/\epsilon k$ is the 2D Fourier
transform of the difference of the in-plane and interplane
Coulomb interaction, and $\nu=m/\pi\hbar^2$ is the   compressibility.

\noindent{\it Discussion of Eq.(\ref{Gap}).}\hskip4mm
We can evaluate the contributions of two regions in the
$k$-space,
$kR_c \gg 1$ and $kR_c \ll 1$, by replacing
$J_0(x)$ with its asymptotic expressions:
  \begin{eqnarray}
\label{capacitor}
{\rm (i)}\ \Delta_{kR_c \gg 1}  &=&  \frac{e^2}{\pi\epsilon\hbar
v_F}\hbar\omega_c \ln \frac{d}{r_s};\\
\label{ABG}
\Delta_{kR_c \ll 1} &=&  \frac{\hbar\omega_c^2}{m v_F^2} \ln\frac{d}{r_s}\ .
  \end{eqnarray}
According to the experiment~\cite{WeakFieldGap}, we now assume
that $l_B<r_s<d\ll R_c$, where $l_B$ is magnetic length, and
$r_s$ is the Coulomb screening length. The
contribution (\ref{capacitor}) coincides with the estimate
(\ref{gap}), except for the log term suppressed by screening.
When the result (\ref{capacitor}) is compared to the
experiment~\cite{WeakFieldGap}, one has to be careful because
the screening length is probably set by the well width, rather
than by the electrons' compressibility. Because this may alter the
log term, we can claim only  an agreement with the experimentally
measured $\Delta$ by the order of magnitude.

It is interesting to note that the contribution (\ref{capacitor})
coincides with the gap found by hydrodynamics
theory~\cite{Glazman}, the difference in the log term being due
the effect of screening by another plane. The two terms
(\ref{capacitor}),(\ref{ABG}) account for contributuions of the
two parts of the charge density, the singular and the smooth.
Not surprisingly, the singular density dominates in the energy,
which one can understand by comparing charging energy of a
two-ring capacitor with that of a parallel plate capacitor, both
of the size $R_c$.

\noindent{\it Bosonization calculation: a sketch.}\hskip4mm
Now we proceed with an accurate derivation of tunneling current.
In the weak field, $\hbar\omega_c\ll E_F$,
the electron motion is semiclassical. Therefore,  instead  of  doing full
many-body  theory,  we  can write down the action that corresponds to
the classical Landau Fermi liquid equation, and use it to  study
dynamics  in  imaginary  time. By that, we will determine optimal
path in imaginary time (instanton) whose action  will  give  the
exponent of the tunneling rate.

The advantage of this approach, besides making a clear relation
with classical theory, is the possibility of including effects of a  parallel
magnetic field and of a scattering by disorder in a natural fashion.

We use a  bosonized Fermi liquid
picture~\cite{bosonization,bosonization:application}
  %\cite{Luther,Haldane,Fradkin,Houghton}
to write electron operators in terms of Bose fields:
   \be
\label{Psi}
\psi_{R(L)}({\bf r})=
\frac{1}{\sqrt{2\pi v_F \tau_0}}
\int e^{i p_F {\bf  n}{\bf  r} - i \phi_{R(L)}({\bf n},{\bf r},t)}
d{\bf n}\ .
  \ee
Here the unit vector ${\bf n}={\bf p}/p_F$ labels points of the
2D Fermi-surface, and $\tau_0$ is a cutoff of the order of
the inverse bandwidth that appears in the bosonization formalism.
The integral $\int...d{\bf n}$ means $\int...d\theta/2\pi$,
where $\theta$ is the polar angle in the ${\bf p}-$plane.

The bosonized imaginary time action for $\phi({\bf n},{\bf r},t)$ is
   \be
\label{action}
S = \frac{m}{4\pi}\int  \phi\,D
\left( i \partial_t - D \left(  1 + \hat{F}\right)\right)
\phi\,d{\bf n}d^2{\bf r}\,dt\ .
   \ee
Here $\phi = (\phi_R, \phi_L)$, and $\hat{F}$ is the integral
operator (\ref{interaction}).

Note that $u=D\phi({\bf n},{\bf r},t)$ has a clear meaning of
the displacement of the Fermi-surface at the point ${\bf n}$ in
the normal direction, i.e., along ${\bf n}$. The saddle points
$\phi({\bf n},{\bf r},t)$ of this action satisfy
Eq.(\ref{LandauFermi}). According to Haldane~\cite{bosonization}, in
the context of the Fermi liquid bosonization, the operator $D$
has an interpretation of a covariant derivative.

To get the tunneling current, one has to find the amplitude
  \be
K({\bf r},\tau) = \langle {\rm T}_\tau\psi^{+}_R({\bf r},i\tau) \psi_R(0,0)
\psi_{L}({\bf r},i\tau)
\psi_L^{+}(0,0)\rangle\ .
   \ee
Then, one continues $K$ to the real time axes from the upper and
the lower halfplane. (Following~\cite{Kadanoff}, we denote these
functions as $K^{>(<)}({\bf r},\tau)$. By the standard
formalism~\cite{Mahan}, one can express the tunneling
current as
   \be\label{current}
I = e |t_0|^2 {\bf Im} \int d^2x\,dt
\left(
\Phi_{{\bf r},t}e^{-ieVt}
+ \Phi_{-{\bf r},-t}e^{ieVt}
\right)\ ,
   \ee
where $\Phi_{{\bf r},t}= K^{>}({\bf r},t) - K^{<}({\bf r},t)$. To
find $K$, one has to evaluate the functional integral of the
product of the exponentials (\ref{Psi}) with the weight
$e^{-S}$. One notes that the integral is gaussian, and writes:
  \be\label{amplitudeK}
K({\bf r},t)=\frac{1}{(2\pi v_F\tau_0)^2}
\int\, e^{-S_{{\bf n}_0,{\bf n}_1}}
d{\bf n}_0\,d{\bf n}_1 \ .
  \ee
Here the action
  \be
\label{saddle-point}
S_{{\bf n}_0,{\bf n}_1}
= \frac{4\pi^2}{m} \langle
J
D^{-1}\left(i\partial_t - D(1+\hat{F})\right)^{-1}
J \rangle\ ,
  \ee
where $\langle...\rangle=\int... d{\bf n}\,d^2{\bf R}\,dt$, and
  \be\label{source}
J =
\delta({\bf r})\delta(t)\delta({\bf n}\!-\!{\bf n}_0)
- -  \delta({\bf r}\!-\!{\bf R}) \delta(t\!-\!\tau)\delta({\bf n}\!-\!{\bf
n}_1).
  \ee
The source term $J$ describes an electron injected at $t=0$, ${\bf r}=0$
at the point ${\bf n}_0$ of the Fermi surface, and then
removed at $t=\tau$, $\bf r=\bf R$ from a different point
${\bf n}_1$.

Calculation of the action (\ref{saddle-point}) involves
inverting the operator $i\partial-D(1+\hat F)$, which has been
done above (see (\ref{ladder}), (\ref{inverse_operator})), and
then applying it to $J$. The latter procedure requires a lot of
attention, because of the singular nature of the operators in
(\ref{saddle-point}). A complete discussion of this operator,
including its relation to classical trajectories,
and regularization technique will be presented
elsewhere~\cite{LS2}. The result of this calculation, however,
is very simple: $I(V)\simeq\delta(eV-\Delta){\rm sign}(V)$,
where $\Delta$ is given by Eq.(\ref{Gap}). In other words, the
rigorous calculation confirms the semiclassical result
(\ref{Gap}) for the gap, and predicts an infinitely sharp
resonance peak at $eV=\Delta$.

To include scattering by disorder, we modify the
action (\ref{action}) by adding a ``collision integral'' term:
  \be\label{ScatteringAction}
S = \frac{m}{4\pi}\int
\phi_{-\omega}({\bf n},{\bf r})
D \hat W
\phi_{\omega}({\bf n},{\bf r})\,
{\rm sign}\omega
\frac{ d\omega }{2\pi}\,
d{\bf n}\,
d^2{\bf r} .
   \ee
Here $\hat W$ is the collision integral operator
\be
\hat W u\ ({\bf n}) =
- -\frac{1}{\tau_s}(u({\bf n})-\int u({\bf n})\,d{\bf n})
\ ,
\ee
taken   in   the   so   called   $\tau-$approximation.   It   is
straightforward  to  verify  that  the  new  action leads to the
classical Fermi liquid  kinetic  equation  with  the  scattering
term.  Let us note that in the imaginary time representation the
action (\ref{ScatteringAction})  is  non-local,  because  it  is
dissipative.

The Gaussian functional integral over $\phi$ again gives
(\ref{amplitudeK}), where now
  \be
\label{saddle-point-W}
S_{{\bf n}_0,{\bf n}_1}
= \frac{4\pi^2}{m} \langle
JD^{-1}\left(i\partial_t -
D(1+\hat{F})-\hat W \right)^{-1}J\rangle\ .
  \ee
In the absence of a  magnetic field, by evaluating the action
(\ref{saddle-point-W}), then plugging it in
Eqs.(\ref{amplitudeK}),(\ref{current}), and doing the
integral~\cite{LS2}, we get
  \be\label{NoFieldIV}
I_0(V) = \frac{e|t_0|^2}{2\pi v_F p_F \tau_s}
\frac{eV}{e^2V^2 + \frac{1}{\tau_s^2}}\ .
  \ee
The peak of conductivity $dI/dV$ is located at zero bias, and
has finite width due to elastic scattering, and Lorentzian
tails, in agreement with the experiment~\cite{B=0peak,WeakFieldGap}.

A perpendicular magnetic field creates the tunneling gap.
Calculation~\cite{LS2} shows that the $I-V$ curve is related
to the zero-field curve (\ref{NoFieldIV}) in a simple way:
  \be\label{FullIV}
I(V) =
\left\{
\begin{array}{rcl}
I_0(V - \Delta/e) &{\rm at}& V>\Delta\ ;\\
0 &{\rm at}& -\Delta<V<\Delta\ ;\\
I_0(V + \Delta/e) &{\rm at}& V<-\Delta
\end{array}
\right.\ ,
   \ee
where  $\Delta$  is  given  by  Eq.(\ref{gap}).  So,  at  finite
$B_\perp$ the peaks of $I_0(V)$ are simply shifted by $\Delta/e$
away from zero bias.

This formula correctly describes the width of the gap and the
peak shape, however it doesn't hold deep inside the gap, near
$V=0$, because our simple approach to scattering breaks down at
$t \gg \tau_s$. In this limit the charge spreading regime
changes to diffusive, and one has to deal with a
classical electrodynamis problem~\cite{LS1}.

To include a  parallel field, we modify the amplitude $K({\bf
r},t)$ by introducing in (\ref{amplitudeK}) the Aharonov-Bohm
phase given by the parallel field flux:
  \be
\tilde K({\bf r},t)=\,
e^{-{ie\over\hbar c}B_\parallel d y}
K({\bf r},t)\ ,
  \ee
where $y$ is the component of the radius vector ${\bf r}$
perpendicular to the field $B_\parallel$. The expression
(\ref{current}) then produces the integral (\ref{ABphase}) equal
to $J_0^2\left(\frac{p_Fd}{\hbar B_\perp} B_\parallel\right)$.
This gives a prefactor to the tunneling current, which makes the
current an oscillating function of $B_\parallel$.

\noindent{\it   Conclusion.}\hskip4mm
We interprete the $I-V$ curve in terms of a resonance tunneling
mechanism that involves an excitonic state, whose structure in
the weak field limit can be studied by the Fermi liquid theory.
  %It consists of an electron and a hole localized near cyclotron
  %orbits, and of a smooth Fermi liquid background distribution.
We calculate the tunneling gap, and find linear dependence on the
magnetic field, in agreement with the experiment.
The structure of the excitonic state
manifests itself in the gap high sensitivity to any asymmetry between
the wells, like density mismatch. Also, it can be
probed by a magnetic field parallel to the barrier,
by observing oscillatory dependence of the current on the field.

We study the problem by means of the Fermi liquid bosonization
theory, and obtain a complete $I-V$ curve that includes the
effects of the perpendicular and parallel fields, as well as
of elastic scattering.

\end{document}